\magnification=1200
\baselineskip=20 pt

\def\lijk{\lambda^k_{ij}}
\def\lpijk{\lambda^{\prime k}_{ij}}
\def\lppijk{\lambda^{\prime \prime k}_{ij}}
\def\dbri{\bar{d}_{Ri}}
\def\ddk{D^{+}_k}
\def\lj{l_j}
\def\lcbi{\bar {l}^c_i}

\def\nucbli{\bar {\nu}^c_{Li}}
\def\ecbli{\bar{e}^c_{Li}}

\hfill{hep-ph/9909518}

\centerline{\bf Neutrino masses and mixing angles from  leptoquark
interactions}
\vskip 1 true in
\centerline{\bf Uma Mahanta}
\centerline{\bf Mehta Research Institute}
\centerline{\bf Chhatnag Road, Jhusi}
\centerline{\bf Allahabad-211019, India}

\vskip .4 true in

\centerline{\bf Abstract}

In this paper we show that the mixing between leptoquarks
(LQ's) from different $SU(2)_l$
multiplets can generate a non-trivial Majorana mass matrix for neutrinos
through one loop self energy diagrams. Such mixing can arise from
gauge invariant and
 renormalizable LQ-Higgs  interaction terms after EW
symmetry breaking. We use the experimental indication on neutrino
oscillation to find constraints on specific combinations of LQ
couplings to quark-lepton pairs and to the SM higgs boson.
These constraints are compared with the ones from
$\pi\rightarrow e\bar {\nu}_e$.


\vfill\eject

\centerline{\bf I. Introduction}

The recent neutrino data from Superkamiokande (SK) [1] has provided
strong evidence for $\nu_{\mu}$ oscillating into $\nu_{\tau}$
or some species of sterile neutrinos as an explanation of the atmospheric
neutrino anomaly. The observed solar neutrino deficit [2] is also probably
an indication of $\nu_e$ oscillating into some other species of neutrinos.
Results from the laboratory experiment by the LSND collaboration [3] can 
also probably be considered as yet another type of neutrino oscillation.
Neutrino oscillation imply a non-trivial structure of the neutrino mass 
matrix.
The extreme smallness of neutrino masses compared to quarks and charged
lepton masses naturally suggests that they are zero at the tree level 
but are generated through higher order loop corrections. The coupling
constants involved in this radiative mass generation must be small
in order to reproduce the small neutrino masses. In this paper we shall
adopt this viewpoint and show how the experimentally suggested neutrino 
masses and mixing angles  could be generated at one loop level through
higgs induced mixing between different leptoquark (LQ) multiplets.

Leptoquarks (LQ's) [4] occur naturally in many extensions of the SM namely 
grand unified models, extended technicolor models and models of quark
lepton substructure which contain quarks and leptons in the same multiplet.
If LQ's couple both to quark pairs and quark-lepton pairs then they
mediate too rapid proton decay. To avoid this disaster and to keep the LQ 
interpretation interesting for low energy phenomenology it is
necessary that LQ coupings to quark pairs must vanish.
In order to enable LQ's to couple to quark-lepton pairs in the context
of the $SU(3)_c\times SU(2)_l\times U(1)_y$ theory, they must be color 
triplets or antitriplets. Further they can form a singlet, doublet or
triplet representation of $SU(2)_l$. Just like quarks and leptons come in
different generations LQ's can also carry a generation index. In this work 
we shall assume that the number of LQ generations is equal to that of 
fermions ($n_g=3$). This happens in the superstring inspired E(6) model 
and also in the E(8)$\times $E(8) $\sigma $ models.
The coupling constant of LQ's to q-l pairs in 
general can depend on the the generation indices of the LQ, the quark
and the lepton. Such a general flavor structure for LQ couplings
and their phenomenological implications were considered  
 considered in ref. [4].
 The existing flavor
changing neutral current data however strongly constrains the flavor
off diagonal couplings of LQ's to q-l pairs. Besides coupling to q-l
pairs, LQ's can also couple to the SM gauge bosons and to the higgs
doublet. Of particular importance for this paper are the Higgs-LQ
trilinear
interaction terms which lead to mixing between LQ's from different
$SU(2)_l$ multiplets after EW symmetry breaking. We shall show that
this mixing among different LQ multiplets gives rise to a non-trivial
lepton number violating
Majorana mass matrix for neutrinos through loop corrections.

The contents of this article are divided into the following subsections.
In Sec. II we present the interaction Lagrangians of singlet, doublet and
triplet LQ's with q-l pairs and with the higgs doublet. In Sec. III 
we derive the neutrino mass matrix that arises from the mixing of doublet
and singlet LQ's. In Sec. IV we show that the mixing between doublet and 
triplet LQ's
can also generate a non-trivial
mass matrix for neutrinos. In Sec. V we use the experimental indication 
on neutrino oscillation to derive some constraints on specific combinations of
LQ couplings and compare them with the existing constraints.
 Finally in Sec. VI we present the main conclusions of our
study.

\centerline{\bf II. Interaction Lagrangians for Leptoquarks}

We shall assume that the light LQ's present in the low energy theory arise
from an underlying high energy theory that breaks down into the 
SM gauge group
at some high energy scale $\Lambda \gg v$.
Here v is the scale associated with electroweak symmetry breaknig.
 To discuss the
low energy LQ phenomenology
in a model independent way we shall construct its Lagrangian based on
invariance under the SM gauge group and renormalizability.
To satisfy the strong constraints [4] of the helicity suppressed decay
$\pi\rightarrow e\bar {\nu}_e$ on LQ's with non chiral couplings
we shall assume that each LQ couples to quarks of a particular
chirality only. Consider three scalar LQ fields D, S and T
with the following $SU(3)_c\times SU(2)_l\times U(1)_y $ 
assignments: $D\sim (3^*, 2, -{1\over 6})$,
$S\sim (3^*, 1, {1\over 3})$ and $T\sim (3^*, 3, {1\over 3})$. 
The general flavor structure of their Yukawa couplings to q-l pairs
are given by

$$\eqalignno{L_1& =[\sum\lijk\lcbi i\tau_2 q_j S_k +\sum\lpijk
\dbri\lj\ddk\cr
& +\sum\lppijk \lcbi\tau_a i\tau_2 q_j T^a_k] +h.c.\cr
&=[ \sum\lijk (\nucbli d_{Lj}-\ecbli u_{Lj})S_k
+\sum\lpijk\dbri (\nu_{Lj}D^{+1}_k+e_{Lj}D^{+2}_k)\cr
&+\sum\lppijk (\nucbli d_{Lj}+\ecbli u_{Lj})T^3_k+
\sum \lppijk \ecbli d_{Lj} T^{+}_k\cr
&-\sum\lppijk \nucbli u_{Lj} T^{-}_k ]+h.c.
&(1)\cr}$$

In the above i,j,k refer to the generation index of the relevant field.
Repeated indices are all summed over. q and l are LH 
quark and lepton fields. $d_R$ is the RH  down quark field.
$\psi^c=C\bar{\psi}^T$ is the charge conjugated fermion field.
$D^{+1}_k$ is the $I_3={1\over 2}$ component of $D^+_k$.
 $T^{+}_k=T^1_k+iT^2_k$ and $T^{-}_k=T^1_k-iT^2_k$.
$T^+_k$, $T^-_k$ and $T^3_k$ have the following weak isospin
$I_3$,  and charge $Q$
assignments: $T^+_k\sim (1, {4\over 3})$, $T^-_k\sim (0, {1\over 3})$
and $T^-_k\sim (-1, -{2\over 3})$. Since $T^a_k$ carries $U(1)_y$
charge $T^a_k \ne (T^a_k)^+$.
 The above interaction terms 
are written in the gauge or interaction basis for the relevant fields.
From the above Lagrangian it follows that $S_i$, $T_i$ and $D_i$
have lepton numbers of -1, -1 and +1 respectively.
Note that since the Yukawa couplings of $D_i$, $S_i$ and $T_i$ are
chiral in nature they cannot generate any neutrino mass through
radiative corrections unless we add new interactions.

Besides the Yukawa couplings of the LQ's, the low energy effective 
Lagrangian will also contain possible higgs-leptoquark interactions.
The Higgs-LQ interaction can be expressed by the following Lagrangian

$$\eqalignno{L_2&=[\sum K_i (D_i^+\phi_c )S_i+\sum K^{\prime }_i
(\phi_c^+\tau_a D_i ) T_i^{+a} +h.c.]\cr
&-\sum\mu^2_{\alpha}(\chi^{+i}_{\alpha}\chi^{i}_{\alpha})
+\sum h^i_{\alpha} (\phi^+\phi )(\chi^{+i}_{\alpha}\chi^{i}_{\alpha})\cr 
&={v+h\over \sqrt {2}}[\sum K_i D^{1*}_iS_i+K_i^{\prime}D^2_iT^-_i+
K^{\prime}_i D^1_i T^3_i+h.c.]\cr
&-\sum\mu^2_{\alpha}(\chi^{+i}_{\alpha}\chi^{i}_{\alpha})
+\sum h^i_{\alpha} (\phi^+\phi )(\chi^{+i}_{\alpha}\chi^{i}_{\alpha} )
&(2)\cr}$$

where $\phi_c=i\tau_2 \phi^*$.
$\chi^i_{\alpha}$ is a collective symbol for all the LQ fields with
$\alpha $ =1, 2, 3 referring to D, S and T respectively. 
If the SM
Higgs doublet $\phi$ and the LQ's belong to the same multiplet
of some higher symmetry group then such interaction terms can naturally
arise when the higher symmetry group breaks down into $SU(3)_c\times
SU(2)_l\times U(1)_y$ at some high energy scale $\Lambda$.
After EW symmetry breaking at the Fermi scale the Lagrangian
$L_2$ will generate
mixing between different LQ multiplets. In the following we shall
consider the mixing between two different LQ multiplets at a time.
However our results will not be very different from the more general
case where all the mixing terms are present provided there is no
accidental cancellation among different contributions.

\centerline{\bf Neutrino mass matrix from doublet-singlet mixing}

It follows from eqn (2) that after EWSB
 the doublet-singlet mixing term can be written as

$$L_3=\sum K_i({v+h\over \sqrt {2}})(D^{1*}_iS_i+S_i^*D^1_i )\eqno(3)$$

The above Lagrangian mixes $D^1_i$ with $S_i$ both of
which carry the same charge and color quantum numbers.
It can be shown that the LQ fields under consideration in the mass
eigenstate basis are given by
$D^{1\prime}_i=\cos\theta_i D^1_i +\sin\theta_i S_i$ and 
$S_i^{\prime}=-\sin\theta_i D^1_i +\cos\theta_i S_i$.
The mixing angle $\theta_i$ is given by $\sin\theta_i={{vK_i\over \sqrt {2}
b_i}\over [1+{v^2K_i^2\over 2b_i^2}]^{1\over 2}}$. Here 
$a_i=m^2_{Di}+m^2_{Si}$
and $b_i=m^2_{Di}-m^2_{Si}$. $m_{Di}$ and $m_{Si}$ are the shifted masses
after EWSB and are given by $m^2_{Di}=\mu^2_{Di}-h^i_D{v^2\over 2}$ and
$m^2_{Si}=\mu^2_{Si}-h^i_S{v^2\over 2}$. The primed fields refer to the mass 
eigenstate basis. Their masses are given by $m^{\prime 2}_{Di}={1\over 2}
[a_i+b_i+{v^2K_i^2\over b_i}]$ and
$m^{\prime 2}_{Si}={1\over 2}[a_i+b_i-{v^2K_i^2\over b_i}]$.
All phenomenological implications of LQ interactions must be derived
 in terms of fields in the mass eigenstate basis. When the interactions of
$D^1_i$ and  $S_i$ with the d-$\nu$ pair are written in terms of mass
eigenstate fields we get

$$\eqalignno{L_4&=\sum\lijk\nucbli d_{Lj}(\sin\theta_kD^{\prime 1}_k
+\cos\theta_kS^{\prime}_k )\cr
&+\lpijk\dbri \nu_{Lj} (\cos\theta_k D^{\prime 1*}_k-\sin\theta_k 
S^{\prime}_k *)&(4)\cr}$$

Note that the mixing between $D^1_i$ and $S_i$ induced by the
the higgs doublet introduces non-chiral couplings for $D^{\prime 1}_i$
and $S^{\prime}_i$ which enables them to generate neutrino masses
through radiative corrections. It can be shown that one loop self energy
diagrams involving the exchange of $D^{\prime 1}_i$ and $S^{\prime}_i$
gives rise to the following Majorana mass matrix for neutrinos
$$M_{ik} \approx{N_c\over 32\pi^2}\sum_{j,l} [ \lambda^{ l}_{ij}(m_d)_j
\lambda^{\prime l}_{jk}+ \lambda^{\prime l}_{ji}(m_d)_j \lambda^l_{kj} ]
 \ln{m^{\prime 2}_{Dl}\over m^{\prime 2}_{Sl}}
\sin\theta_l\cos\theta_l \eqno(5)$$

In the above we have neglected the matrices $D_L$ and $D_R$
that connect the
quark gauge eigenstates $d_L$ and $d_R$ to their respective mass eigenstates.
We would like to note first that the individual diagrams 
for $D^{\prime 1}$ and $S^{\prime}$ exchange are separately
 log divergent. But
the log divergences cancel each other in the sum yielding a finite result.
Second the neutrino mass matrix is symmetric in flavor
space  which is a requirement of its majorana nature.
Third the extension of the SM considered here 
does not include any right handed
neutrino and therefore the see saw mechanism does not play any role 
in generating neutrino masses here.
The smallness of neutrino masses in our case has to follow from
the smallness of $\lambda$, $\lambda^{\prime}$, $K_i$
and the loop suppression factor of ${1\over 16\pi^2}$.

\centerline{\bf IV. Neutrino masses from doublet triplet mixing}

The mixing between doublet and triplet LQ's can also give rise to a
non-trivial Majorana mass matrix for neutrinos. The mixing arises
from the following  LQ-Higgs interaction term 

$$\eqalignno{L_5&=\sum K^{\prime}_n (\phi^+_c\tau_a D_n)T^{a+}_n+h.c.\cr
&={v+h\over \sqrt {2}}[\sum K^{\prime}_n D^1_n T^{3+}_n+ 
\sum K^{\prime}_i D^2_i T^-_i]+h.c.
&(6)\cr}$$
The above Lagrangian implies that $D^1_n$ mixes with $T^3_n$ and
$D^2_n$ mixes with $T^-_n$.
The general flavor structure of the Yukawa couplings of $T^a_n$ to
q-l pairs is given by

$$\eqalignno{L_6&=\sum\lppijk \lcbi\tau_a i\tau_2 q_j T^a_n+h.c.\cr
&=[\sum \lppijk (\nucbli d_{Lj}+\ecbli u_{Lj})T^3_k+..]+h.c.\cr
&=[ \sum \lppijk (\nucbli d_{Lj}+\ecbli u{Lj} )(T^{\prime 3}_k
\cos\theta^{\prime}_k-D^{\prime 1}_k \sin\theta_k^{\prime})+..]+h.c.
&(7)\cr}$$
where $\theta^{\prime}_k$ is the mixing angle between $T^3_k$
and $D^1_k$. Proceeding as in Sec. III it can be shown that one loop
self energy diagrams involving the exchange of $D^{\prime 1}_k$
and $T^{\prime 3}_k$ generates the following  mass matrix
for neutrinos

$$M_{ik}\approx {N_c\over 32\pi^2} \sum_{l,j}
 [ \lambda^{\prime\prime l}_{ij}
(m_d)_j \lambda^{\prime l}_{jk}+\lambda^{\prime l}_{ji} (m_d)_j
 \lambda^{\prime\prime l}_{kj}]
 \ln{m^{\prime ^2}_{Tl}\over 
m^{\prime 2}_{Dl}} \sin\theta^{\prime}_l \cos\theta^{\prime}_l\eqno(8)$$

\centerline{\bf V. Implications of neutrino oscillation data
on LQ couplings}

The experimental data indicating neutrino oscillation can be used to
find constraints on specific combinations of LQ couplings. To be specific
let us consider the doublet-singlet mixing case. In general the neutrino
mass matrix $M_{ik}$ is not symmetric. Since the neutrino mass matrix
given by eqn (5) is symmetric it can be diagonalized by a 3 $\times $ 3
 orthogonal matrix U.
 The matrix U can be completely specified in terms
of three angles $\theta_{12}$, $\theta_{23}$ and $\theta_{13}$. 
It has been shown [5]
that the recent SK data, the CHOOZ data [6] and the solar neutrino data
can be accomodated in a three neutrino oscillation model  at 99\%
C.L. for $\sin\theta_{12}=.63$, $\sin\theta_{23}=.71$,  
$\sin\theta_{13}=.45$, $\delta m^2_{32}=m^2_3-m^2_2=8\times 10^{-4} ev^2$ and
$\delta m^2_{21}=m^2_2-m^2_1=1\times 10^{-4} ev^2$. The neutrino oscillation
 data determines only the mixing angles and the mass squared differences
leaving the overall scale for neutrino masses undetermined.
 If we take $m_1=.01 ev$,
 the elements of the neutrino mass matrix in the flavor basis
will be given by $M_{11}=.015$
ev, $M_{12}=M_{21}=.007$ ev, $M_{13}=M_{31}=.004$ ev, $M_{22}=.020$ ev,
$M_{23}=M_{32}=.008$ ev and $M_{33}=.021$ ev. We would like to emphasize
that the value of $m_1$ cannot be chosen arbitrarily. It must satisfy
the constraint from neutrinoless double beta decay which provides
a bound on $M_{11}$. The present experimental upper bound on the
effective Majorana neutrino mass in the flavor basis
is given by $M_{11} < 0.2- 0.4$ eV at 90\% CL [7].
The range of the upper bound is mainly due to the uncertainty 
in the theoretical calculation of the nuclear matrix elements. 
 Besides the resulting values of $m_2$
and $m_3$ must satisfy the bounds $m_2<.17$ Mev and $m_3<18$ Mev [8].
Using the expression of $M_{ik}$ derived earlier we can now find constraints
on particular combinations of LQ couplings. Since the values 
of $M_{ik}$ for different i,k do not differ much among them we shall
consider the constraints that arise from only one of them namely $M_{11}$.
We shall assume that the value of $\ln{m^{\prime 2}_{Dl}\over 
m^{\prime 2}_{Sl}}$ do not change appreciably with generation. For 
$m^{\prime}_D$=300 Gev and $m^{\prime}_S$=200 Gev we get
$\sum (\lambda^l_{1j})^2 m_{dj} \sin\theta_l\cos\theta_l\approx .74\times
10^{-9}$ Gev.
 Further if we assume that only one product coupling is
non-zero at a time we get $(\lambda^l_{11})^2\sin\theta_l\cos\theta_l
\approx .74\times 10^{-7}$, $(\lambda^l_{12})^2\sin\theta_l\cos\theta_l
\approx 5.36\times 10^{-9}$ and
 $(\lambda^l_{13})^2\sin\theta_l\cos\theta_l
\approx 1.48\times 10^{-10}$ indicating a flavor dependent heirarchy
in LQ couplings. If we set $M_{11}=.4$ eV which is the present
upper bound from neutrinoless double beta decay
we get instead $(\lambda^1_{11})^2\sin\theta_l\cos\theta_l\approx
1.97\times 10^{-6}$.
 For a LQ mass of 200 Gev, HERA and Tevatron
will soon be able to probe flavor diagonal LQ coulpings for first generation
down to .1. If we set $\lambda^1_{11}\approx .1$ we get $K_1\approx
2.85$ Mev which is close to the light quark masses.
 It is interesting to compare the constraints on LQ couplings 
derived from   neutrino oscillation data with those derived from other low
 energy experiments e.g. $\pi\rightarrow e\bar {\nu}_e$.
The mixing between different multiplets of LQ's introduces non-chiral
couplings for LQ's. Such non-chiral LQ couplings can give rise to 
helicity unsuppressed contributions to $\pi\rightarrow \bar{\nu}_e$ [8].
For the exchange of S  we have $g_R=\lambda^ 1_{11} \sin\theta_1 $
 and
$g_L=\lambda^ 1_{11} \cos\theta_1 $. From ref[4] we then get
$g_l g_R=(\lambda^{\prime 1}_{11})^2\sin\theta_1\cos\theta_1<{m^2_S\over
(100\  Tev)^2}\approx 4\times 10^{-6}$. Actually the contributions due 
to both $D^1_i$ and $S_i$ exchange must be taken into account which raises
the bound to $8\times 10^{-6}$.
 This bound is nearly two orders
of magnitude greater than the value provided by the neutrino oscillation
data for $M_{11}=.015$ eV. However it is close to the value provided
by the neutrino oscillation data for $M_{11}\sim 0.4$ eV.
The next generation of neutrinoless double beta decay is expected to 
be sensitive to values of $M_{11}$ in the range $10^{-2}-10^{-1}$ eV.
It will enable us to decide whether the neutrino data
gives more stringent bounds on LQ couplings than the pion decay rate.

Coming next to other bounds on flavor off-diagonal LQ couplings,
we note that flavor changing radiative deacys like $\mu\rightarrow
e\gamma$ require non-chiral couplings for LQ's and hence mixing between
different multiplets of LQ's. To be specific consider the higgs
induced mixing between $D^1_i$ and $S_i$. But this mixing does not
contribute to $\mu\rightarrow e\gamma$ since $D^1_i$ does not couple to
$e_i-q$ pair. Similarly the mixing between $D^1_i$ and $T^3_i$
or $D^2_i$ and $T^-_i$ does not contribute to radiative muon deacy
because $D^1_i$ and $T^-_i$ does not couple to charged lepton-quark pair.
The decay $\mu\rightarrow e\gamma$ therefore does not impose any bound on 
the parameters of our model.
We would also like to note that the LQ masses chosen by us are
consistent with the latest bounds from HERA and Tevatron [9].
The HERA bounds depend on the value of $g_{lq}$. For $g_{lq}=e$
the HERA bounds are 237 Gev for first generation and 73 Gev for second 
generation. The Tevatron bounds are most stringent for first generation
($>225$ Gev) and become progressively weaker for second ($>131$ Gev)
and third generation ($>95$ Gev). The Tevatron and HERA bounds
however depend on several assumptions the most crucial of which is
the branching ratio of the LQ into e-q or $\nu$-q pair. The bounds get
significantly relaxed
 if the branching ratios are lower than those usually assumed.
Such reduction in branching ratios can take place in models where
there is an appreciable mixing between LQ's from different $SU(2)_l$
multiplets.
  
\centerline{\bf VII. Conclusion }

In conclusion in this paper we have shown that the interaction between 
LQ's and the the SM higgs doublet induces mixing between
 LQ's from different $SU(2)_l$ multiplets after EW symmetry breaking.
  This mixing introduces small
non-chiral couplings for LQ's in the mass eigenstate basis
and generates a non-trivial Majorana
mass matrix for neutrinos through one loop radiative correction.
We have determined the neutrino mass matrix in terms of LQ
couplings and masses
for the singlet-doublet mixing and the doublet-triplet mixing.
Using the recent combined experimental data indicating neutrino 
oscillation data
we have derived constraints on specific combinations of LQ couplings
to q-l pairs and to the higgs boson. Such constraints are comparable
with those derived from $\pi\rightarrow e\bar {\nu}_e$ if the value of
$M_{11}$ is close to the present upper bound provided
by the neutrinoless double beta decay.

\centerline{\bf References}

\item{1.} Y. Fukuda et al., Phys. Rev. Lett. 81, 1562 (1998);
hep-ex/9805006, hep-ex/9805021 and hep-ex/9807003.

\item{2.} B. T. Cleveland et al., Nucl. Phys. B (Proc. Suppl.) 38,
47 (1995); Kamiokande Collaboration, Y. Fukuda et al., Phys. Rev. 
Lett. 77, 1683 (1996); GALLEX Collaboration, W. Hampel et al.,
Phys. Lett. B 388, 384 (1996); SAGE Collaboration, J. N. Abduarashitov
et al., Phys. Rev. Lett. 77, 4708, (1996); J. N. Bahcall, P. I. Kratev
and A. Yu. Smirnov, hep-ph/9807216.

\item{3.} LSND Collaboration, C. Athanassopoulos et al., Phys.
Rev. Lett. 75, 2560 (1995); 77, 3082 (1996).

\item{4.} W. Buchmuller, Acta. Phys. Austriaca, Suppl 27, 517 (1985);
W. Buchmuller and D. Wyler, Phys. Lett. B, 177, 377 (1986);
W. Buchmuller, R. Ruckl and D. Wyler Phys. Lett. B 191, 442 (1987).

\item{5.} G. L. Fogli, E. Lisi, A. Marrone and G. Sciascia,
Phys. Rev. D 59, 033001 (1999).

\item{6.} CHOOZ Collaboration, M. Apolonio et al., Phys. Lett. B
420, 397 (1998).

\item{7.} L. Baudis et al., Phys. Rev. Lett. 83, 41 (1999), 
hep-ex-9902014.

\item{8.} M. Hirsch et al.,
Phys. Lett. B 378, 17 (1996); Phys. Rev. D 54, 4207 (1996).

\item{9.} Review of Particle Properties, Euro. Phys. Jour. C 3, 1 (1998).

\end